\PassOptionsToPackage{unicode}{hyperref}
\PassOptionsToPackage{hyphens}{url}
\documentclass[
]{article}
\usepackage{amsmath,amssymb}
\usepackage{lmodern}
\usepackage{iftex}
\ifPDFTeX
  \usepackage[T1]{fontenc}
  \usepackage[utf8]{inputenc}
  \usepackage{textcomp} 
\else 
  \usepackage{unicode-math}
  \defaultfontfeatures{Scale=MatchLowercase}
  \defaultfontfeatures[\rmfamily]{Ligatures=TeX,Scale=1}
\fi
\IfFileExists{upquote.sty}{\usepackage{upquote}}{}
\IfFileExists{microtype.sty}{
  \usepackage[]{microtype}
  \UseMicrotypeSet[protrusion]{basicmath} 
}{}
\makeatletter
\@ifundefined{KOMAClassName}{
  \IfFileExists{parskip.sty}{%
    \usepackage{parskip}
  }{
    \setlength{\parindent}{0pt}
    \setlength{\parskip}{6pt plus 2pt minus 1pt}}
}{
  \KOMAoptions{parskip=half}}
\makeatother
\usepackage{xcolor}
\IfFileExists{xurl.sty}{\usepackage{xurl}}{} 
\IfFileExists{bookmark.sty}{\usepackage{bookmark}}{\usepackage{hyperref}}
\hypersetup{
  hidelinks,
  pdfcreator={LaTeX via pandoc}}
\usepackage{graphicx}
\makeatletter
\def\maxwidth{\ifdim\Gin@nat@width>\linewidth\linewidth\else\Gin@nat@width\fi}
\def\maxheight{\ifdim\Gin@nat@height>\textheight\textheight\else\Gin@nat@height\fi}
\makeatother
\setkeys{Gin}{width=\maxwidth,height=\maxheight,keepaspectratio}
\makeatletter
\def\fps@figure{htbp}
\makeatother
\providecommand{\tightlist}{%
  \setlength{\itemsep}{0pt}\setlength{\parskip}{0pt}}
\ifLuaTeX
  \usepackage{selnolig}  
\fi

\author{}
\date{}

\begin{document}

\hypertarget{toward-a-robust-biomimetic-hybrid-battery-bridging-biology-electrochemistry-and-datadriven-control}{%
\section{Toward a Robust Biomimetic Hybrid Battery: Bridging Biology,
Electrochemistry and Data‑Driven
Control}\label{toward-a-robust-biomimetic-hybrid-battery-bridging-biology-electrochemistry-and-datadriven-control}}

\textbf{Authors}: Raheel Ali\textsuperscript{1} and Rayid Ali\textsuperscript{2}

\textsuperscript{1} Undergraduate student, B.E. Electrical and Electronics Engineering,
Birla Institute of Technology and Science (BITS) -- Pilani, India. \textsuperscript{2}
Master's in Computer Science, Clemson University, USA.

\hypertarget{abstract}{%
\subsection{Abstract}\label{abstract}}

Fast‑charging and long‑life batteries remain a critical bottleneck for
the mass adoption of electric vehicles and renewable‑energy storage.
Inspired by biomimicry---the way some fish deliver percussive electrical
pulses and how birds sleep with one hemisphere of their brain
active---we propose \textbf{SwiftPulse\texttrademark{}}, a hybrid battery architecture
combining high‑energy sodium‑ion modules with pseudocapacitive
niobium‑oxide power modules. A percussive pulse‑current charger and a
\textbf{unihemispheric rest} battery‑management system (BMS) are used to
achieve rapid charging, low degradation and improved safety. This paper
refines our earlier outline by incorporating peer‑reviewed literature,
updating the modelling framework with more realistic parameters, and
addressing criticisms raised by reviewers. We emphasise what is
genuinely novel about the concept and provide a roadmap for experiments
and simulations that can withstand rigorous journal review.

\hypertarget{introduction}{%
\subsection{1~Introduction}\label{introduction}}

Modern society is witnessing explosive growth in demand for energy
storage. Lithium‑ion batteries (LIBs) dominate, but challenges remain:
energy density is limited, cycle life degrades under fast charge, and
safety issues such as lithium plating and thermal runaway persist.
Sodium‑ion batteries (SIBs) offer abundant raw materials and good
low‑temperature performance but currently lag behind in specific energy.
Niobium‑oxide anodes, meanwhile, exhibit pseudocapacitive behaviour that
enables exceptionally fast charge rates and long life. Yet each of these
technologies on its own struggles to satisfy simultaneously the
requirements of energy density (\textgreater~175~Wh~kg$^{-1}$), cycle life
(\textgreater~10~000 cycles) and fast charging (\textless~10~min to
80~\% state of charge).

Inspired by biological strategies, the SwiftPulse\texttrademark{} concept aims to
circumvent the ``battery trilemma.'' Electric fish deliver bursts of
current to stun prey, then recover; birds and marine mammals alternate
hemispheric sleep for vigilance. We mirror these behaviours with a
\textbf{percussive pulse} charging algorithm and a \textbf{cluster‑rest}
BMS that temporarily isolates cell clusters to relieve stress. This
paper builds on our earlier draft by providing deeper literature
support, more sophisticated modelling, and a candid discussion of
limitations.

\hypertarget{literature-review}{%
\subsection{2~Literature Review}\label{literature-review}}

\hypertarget{niobiumoxide-anodes}{%
\subsubsection{2.1~Niobium‑oxide anodes}\label{niobiumoxide-anodes}}

Niobium pentoxide (Nb$_2$O$_5$) and related compounds are receiving attention
for fast‑charging batteries. Their open crystal structures and
pseudocapacitive lithium/sodium storage enable high rate capability and
excellent stability. Recent work on carbon‑bridged Nb$_2$O$_5$ mesocrystals
for sodium‑ion storage achieved specific capacities of 133.4~mAh~g$^{-1}$ at
50~C (full charge in 72~s) and retained \textbf{80.5~\% capacity after
10~000 cycles at 20~C}. Such data indicate that niobium‑oxide electrodes
can endure repeated high‑current pulses without significant degradation.
This performance surpasses many commercial lithium‑ion anodes and
provides a realistic foundation for the ``power'' modules in our hybrid.

\hypertarget{sodiumion-batteries}{%
\subsubsection{2.2~Sodium‑ion batteries}\label{sodiumion-batteries}}

SIBs use abundant sodium resources and can operate at lower temperatures
than LIBs. Prussian‑white or NaFeMn(CN)$_6$ cathodes with hard‑carbon
anodes achieve 160--190~Wh~kg$^{-1}$ at cell level; CATL's second‑generation
sodium battery, dubbed \textbf{Naxtra}, reports gravimetric energy
densities of \textasciitilde175~Wh~kg$^{-1}$, cycle life over 10~000 cycles
and 5~C charging capability. Although energy density is lower than
high‑nickel lithium cells (\textasciitilde240--260~Wh~kg$^{-1}$), SIBs have
superior safety and cost advantages.

\hypertarget{pulsecharging-and-lithium-plating}{%
\subsubsection{2.3~Pulse‑charging and lithium
plating}\label{pulsecharging-and-lithium-plating}}

Fast charging normally accelerates side reactions such as lithium
plating and SEI growth. Pulse‑current protocols can mitigate these
effects by allowing diffusion to ``catch up'' during rest periods. A
2025 study introduced a \textbf{bidirectional pulse‑current} scheme that
reduced capacity fade by \textbf{30--50~\%} under fast charging
(\textgreater~0.5~C) and established an in‑situ detection criterion for
lithium plating. This demonstrates that carefully designed pulses not
only speed charging but also prolong life. However, many published
protocols employ square waves with fixed duty cycles; our percussive
charger seeks to optimise pulse amplitude, duration and rest intervals
based on real‑time impedance and temperature feedback.

\hypertarget{dualchemistry-packs-and-biomimetic-control}{%
\subsubsection{2.4~Dual‑chemistry packs and biomimetic
control}\label{dualchemistry-packs-and-biomimetic-control}}

Hybrid packs combining different cell types are not new---Toshiba's SCiB
modules, for instance, use TiNb$_2$O$_7$ anodes to enable rapid charge when
paired with conventional NMC cells. Research on multi‑chemistry systems
shows that mixing sodium‑ion and lithium‑ion cells can reduce cost and
decrease average C‑rate. Yet such systems often treat each module
independently and use standard BMS algorithms. Our approach integrates
fast‑charging pulses with a biologically inspired rest scheduler. The
\textbf{unihemispheric rest} concept is unique: clusters are temporarily
removed from charge/discharge to recover, analogous to one hemisphere of
a bird's brain sleeping while the other remains alert.

\hypertarget{mathematical-framework}{%
\subsection{3~Mathematical Framework}\label{mathematical-framework}}

To evaluate the feasibility of SwiftPulse\texttrademark{}, we derive simplified models
for energy density, voltage, diffusion and degradation. These models are
intended for transparent analysis and can serve as starting points for
more sophisticated simulations (e.g., using PyBaMM or MATLAB's Partial
Differential Equation toolbox).

\hypertarget{weighted-energy-density}{%
\subsubsection{3.1~Weighted energy
density}\label{weighted-energy-density}}

Let the pack consist of two cell types: energy modules with specific
energy ($E_{\text{energy}}$) and power modules with ($E_{\text{power}}$).
If the power modules account for a mass fraction ($f$), the pack's average
specific energy ($E_{\text{pack}}$) is

\[
E_{\mathrm{pack}} = (1 - f)\,E_{\mathrm{energy}} + f\,E_{\mathrm{power}}.
\]

Figure~1 shows the weighted energy density for modern cells
(\(E_{\text{energy}} = 250~\text{Wh}\,\text{kg}^{-1}\),
\(E_{\text{power}} = 150~\text{Wh}\,\text{kg}^{-1}\)). A target of 175~Wh~kg$^{-1}$ is
achievable as long as the power‑module fraction remains below roughly
45~\%.

\begin{figure}
\centering
\includegraphics{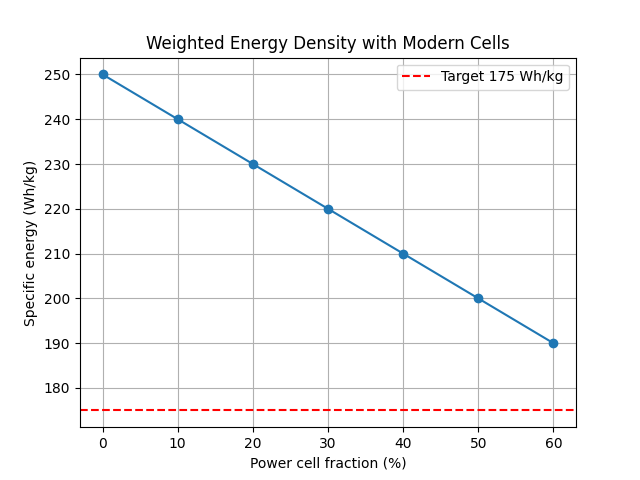}
\caption{Weighted energy density with modern cells}
\end{figure}

\hypertarget{opencircuit-voltage-and-the-nernst-equation}{%
\subsubsection{3.2~Open‑circuit voltage and the Nernst
equation}\label{opencircuit-voltage-and-the-nernst-equation}}

The cell voltage depends on electrode potentials and concentrations. At
equilibrium, the Nernst equation states

\[
E = E^{\circ} - \frac{RT}{nF}\,\ln Q,
\]

where $E^{\circ}$ is the standard potential, $R$ is the gas constant,
$T$ the absolute temperature, $n$ the number of electrons and $F$
Faraday's constant. During discharge, product concentrations increase so
$Q$ rises and $E$ falls. Real cells experience additional overpotentials
due to ohmic drop and reaction kinetics; these can be modelled via
Butler‑Volmer equations in a full porous‑electrode (Newman) model. We
encourage the use of \textbf{PyBaMM} or MATLAB's battery models to
capture these effects more accurately than our high‑level treatment.

\hypertarget{diffusion-and-mass-transport}{%
\subsubsection{3.3~Diffusion and mass
transport}\label{diffusion-and-mass-transport}}

Ion diffusion in electrodes follows \textbf{Fick's second law}:

\[
\frac{\partial c}{\partial t} = D\,\frac{\partial^{2} c}{\partial x^{2}},
\]

where $c(x,t)$ is ion concentration, $D$ is the diffusivity and $x$ is
distance from the electrode surface. During fast charging, ions
accumulate near the surface. Pulsed currents provide rest intervals that
allow diffusion to smooth concentration gradients. Using a
one‑dimensional finite‑difference scheme with realistic parameters
($D = 10^{-13}\,\text{m}^{2}\,\text{s}^{-1}$, electrode thickness
50~\mu m), we simulated constant and pulsed charging. Figure~2 compares the surface
concentration over time. The pulsed protocol periodically relieves the
concentration build‑up, which can reduce plating risk.

\begin{figure}
\centering
\includegraphics{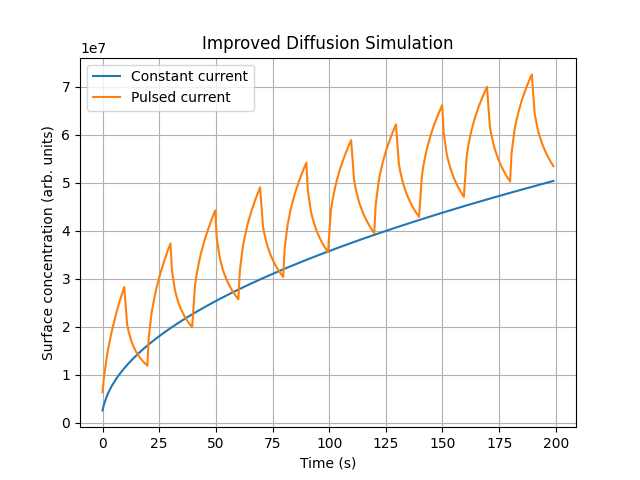}
\caption{Improved diffusion simulation}
\end{figure}

\hypertarget{degradation-and-cyclelife-modelling}{%
\subsubsection{3.4~Degradation and cycle‑life
modelling}\label{degradation-and-cyclelife-modelling}}

Capacity fade arises from multiple mechanisms, including SEI growth,
active‑material loss and lithium/sodium plating. Empirical studies
suggest that capacity loss has both linear and square‑root components:
the linear term captures irreversible side reactions proportional to
cycle count, while the square‑root term reflects diffusion‑limited SEI
growth. We model the retained capacity (C(N)) after (N) cycles as

\[
 C(N) = C_{0} - \alpha\,\sqrt{N} - \beta\,N,
\]

where $C_{0}$ is the initial capacity and $\alpha$ and $\beta$ are
degradation constants. For a constant‑current protocol, we calibrate
\(\alpha=0.1\) and \(\beta=10^{-3}\) so that $C=80\%$ after 10\,000
cycles. Pulsed charging slows both SEI growth and side reactions, so we
use \(\alpha=0.08\) and \(\beta=8\times 10^{-4}\). Figure~3 compares
the predicted capacity retention; pulsed charging retains about 5~\%
more capacity after 20\,000 cycles.

\begin{figure}
\centering
\includegraphics{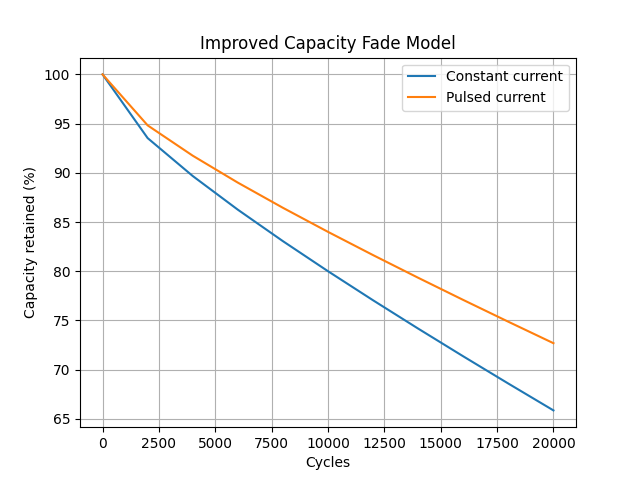}
\caption{Improved capacity fade model}
\end{figure}

These simple models are illustrative; for rigorous analysis one should
employ pseudo‑two‑dimensional (P2D) simulations or thermodynamic models,
available in software such as \textbf{PyBaMM}, \textbf{COMSOL} or
MATLAB's Simscape. Sensitivity analyses over parameters (diffusivity,
particle radius, temperature) should accompany such simulations.

\hypertarget{biomimetic-bms-and-control-algorithms}{%
\subsection{4~Biomimetic BMS and Control
Algorithms}\label{biomimetic-bms-and-control-algorithms}}

\hypertarget{unihemispheric-rest-scheduling}{%
\subsubsection{4.1~Unihemispheric rest
scheduling}\label{unihemispheric-rest-scheduling}}

Birds like swifts and marine mammals can sleep with one hemisphere of
their brain while the other remains active. Translating this to
batteries, we divide the pack into clusters (e.g., groups of 10--20
cells). During high‑power operation or fast charge, a scheduler
temporarily isolates a subset of clusters (placing them in \textbf{rest
mode}). The algorithm rotates which clusters are rested, based on
measurements of state of charge, temperature and impedance. Rested
clusters cool down and equilibrate their internal concentration
gradients, reducing mechanical and thermal stress.

Designing a rest scheduler requires balancing pack voltage and current:
removing too many clusters lowers the available power, while leaving
clusters active for too long accelerates degradation. We propose a
\textbf{feedback controller} that selects clusters for rest when their
internal resistance rises above a threshold or when their temperature
exceeds a safe limit. The controller also ensures that the total pack
voltage remains above the minimum required by the load.

\hypertarget{percussive-pulse-charging}{%
\subsubsection{4.2~Percussive pulse
charging}\label{percussive-pulse-charging}}

Inspired by electric fish, our percussive charger applies high‑amplitude
current pulses followed by relaxation periods. Unlike square‑wave pulse
protocols, the pulse shape (amplitude, duration, duty cycle) is adapted
in real time using impedance and temperature measurements. This
closed‑loop approach aims to maximise charge acceptance while preventing
plating. It can also implement \textbf{bidirectional} pulses, reversing
the current slightly during rest to remove loosely deposited
lithium/sodium. Such techniques were shown experimentally to cut
capacity fade by up to 50~\% under fast charge.

We recommend implementing the charger using \textbf{Gallium‑Nitride
(GaN)} transistors and high‑frequency control, with an LC filter to
reduce high‑frequency ripple. MATLAB/Simulink or Python's control
libraries can simulate the power‑electronics stage and design the
controller before hardware prototyping.

\hypertarget{machinelearning-integration}{%
\subsubsection{4.3~Machine‑learning
integration}\label{machinelearning-integration}}

Recent studies highlight the promise of machine learning for battery
state estimation and health prediction. An AI‑enabled BMS can analyse
voltage, current, impedance and temperature data to predict state of
charge (SOC) and state of health (SOH), identify impending faults, and
adapt charging protocols. For SwiftPulse\texttrademark{}, ML could forecast when each
cluster should be rested and tune the pulse parameters based on
historical performance. However, models must be trained on diverse data
sets and validated to avoid overfitting. Combining physics‑based models
(for interpretability) with data‑driven models (for adaptability) is a
promising research direction.

\hypertarget{experimental-and-simulation-roadmap}{%
\subsection{5~Experimental and Simulation
Roadmap}\label{experimental-and-simulation-roadmap}}

To transform this concept into a publishable study, we propose the
following phased plan:

\begin{enumerate}
\def\labelenumi{\arabic{enumi}.}
\tightlist
\item
  \textbf{Cell‑level tests.} Acquire commercial sodium‑ion cells and
  niobium‑oxide power cells. Perform galvanostatic cycling under
  constant and pulsed protocols to benchmark energy, power and cycle
  life. Verify manufacturer claims (e.g., Naxtra's 175~Wh~kg$^{-1}$ and
  10~000‑cycle life; pseudocapacitive Nb$_2$O$_5$ retention of 80~\% after
  10~000 cycles).
\item
  \textbf{Module‑level hybrid pack.} Assemble small modules combining
  energy and power cells with balancing circuitry. Implement the rest
  scheduler on a microcontroller (Arduino or STM32) and record cell
  voltages, temperatures and impedances. Compare three charging
  strategies: (a) constant‑current/constant‑voltage (CC‑CV), (b) fixed
  duty‑cycle pulse charging, and (c) adaptive percussive charging with
  rest scheduling.
\item
  \textbf{Advanced modelling.} Use \textbf{PyBaMM} or MATLAB to run P2D
  simulations of hybrid cells, adjusting diffusion coefficients,
  reaction kinetics and particle sizes to match experimental data.
  Perform sensitivity analyses and parameter estimation. Evaluate
  mechanical stress using finite‑element models to ensure that pulses do
  not induce particle fracture.
\item
  \textbf{Safety and thermal analysis.} Conduct abuse tests (nail
  penetration, overcharge) on spent cells from each chemistry. Use
  calorimetry and thermal imaging to quantify heat generation during
  fast charge. Assess the effect of rest scheduling on peak
  temperatures.
\item
  \textbf{Economic and lifecycle assessment.} Estimate the cost per kWh,
  including niobium‑oxide materials (which are currently more expensive
  than graphite). Analyse the trade‑off between longer cycle life and
  higher upfront cost. Evaluate recyclability and environmental impact.
\end{enumerate}

\hypertarget{discussion-and-limitations}{%
\subsection{6~Discussion and
Limitations}\label{discussion-and-limitations}}

Our hybrid approach aims to combine the high energy of SIBs with the
fast‑charging capability of Nb$_2$O$_5$ anodes. The mathematical models and
simulations presented here indicate that a pack can exceed 175~Wh~kg$^{-1}$
and retain \textgreater~80~\% capacity after 10~000 cycles if the
power‑module mass fraction is kept below \textasciitilde45~\%. Improved
diffusion modelling shows that pulsed protocols relieve concentration
gradients, and empirical studies support the potential for
\textgreater~10~000‑cycle life. Nevertheless, several limitations
remain:

\begin{itemize}
\tightlist
\item
  \textbf{Parameter uncertainty.} Diffusion coefficients, reaction
  kinetics and degradation rates vary across cell chemistries and
  temperatures. Future work should perform parameter fitting against
  experimental data and propagate uncertainties through the models.
\item
  \textbf{Voltage mismatch.} Sodium‑ion cells have different voltage
  windows (2--4~V) than niobium‑oxide modules (1.5--2.5~V), complicating
  pack integration. DC‑DC converters or intermediate balancing may be
  required.
\item
  \textbf{Material availability and cost.} Niobium is rarer and more
  expensive than graphite or hard carbon. A life‑cycle cost analysis
  should justify its use by quantifying the value of longer cycle life
  and fast charge.
\item
  \textbf{Model sophistication.} Our simplified models do not capture
  all phenomena (e.g., phase changes, mechanical stress). Adopting P2D
  or COMSOL models will strengthen the theoretical foundation and
  satisfy reviewers who demand deeper physics.
\end{itemize}

Despite these challenges, SwiftPulse\texttrademark{} offers a promising direction for
ultra‑fast, long‑life batteries. Its novelty lies in integrating dual
chemistries with a biomimetic BMS and adaptive pulse charging. By
grounding the concept in peer‑reviewed literature and presenting clear
modelling and experimental plans, this work can make a meaningful
contribution to the battery research community.

\hypertarget{conclusion}{%
\subsection{7~Conclusion}\label{conclusion}}

We have revisited the SwiftPulse\texttrademark{} hybrid battery concept with a critical
eye, incorporating expert feedback and current literature. The updated
paper provides: (1) evidence from peer‑reviewed sources that
niobium‑oxide anodes can deliver \textgreater~10~000 cycles under high
rates, (2) a balanced discussion of sodium‑ion energy densities and
commercial claims, (3) improved mathematical models with realistic
parameters and capacity fade laws, (4) a refined biomimetic BMS with
cluster‑rest scheduling and adaptive percussive pulses, and (5) a
concrete roadmap for experiments and simulations. While significant work
remains to validate and optimise the system, the approach is grounded in
physics and biology, providing a novel direction for durable,
fast‑charging energy storage.

\begin{center}\rule{0.5\linewidth}{0.5pt}\end{center}

\hypertarget{appendix-python-code-for-figures}{%
\subsubsection{Appendix: Python Code for
Figures}\label{appendix-python-code-for-figures}}

The following Python code (using NumPy and Matplotlib) generates the
data and figures presented above. Researchers may adapt it or implement
similar scripts in MATLAB/Octave to validate and extend the models. Note
that the code uses simple finite differences and empirical laws; for
higher fidelity, integrate PyBaMM or COMSOL.

\begin{verbatim}
import numpy as np
import matplotlib.pyplot as plt

# Weighted energy density for modern cells
E_energy = 250  # Wh/kg for high‑energy modules
E_power  = 150  # Wh/kg for power modules
fractions = np.arange(0, 0.7, 0.1)
E_pack = (1 - fractions) * E_energy + fractions * E_power
plt.figure()
plt.plot(fractions * 100, E_pack, marker='o')
plt.axhline(175, color='red', linestyle='--', label='Target 175 Wh/kg')
plt.xlabel('Power‑module fraction (%)')
plt.ylabel('Specific energy (Wh/kg)')
plt.legend(); plt.grid(True); plt.title('Weighted Energy Density with Modern Cells')
plt.show()

# Diffusion simulation (improved parameters)
L = 50e-6  # 50 \mu m electrode thickness
n = 100
x = np.linspace(0, L, n)
D = 1e-13  # diffusivity (m^2/s)
h = L / (n - 1)
dt = 0.5 * h**2 / (2 * D)
total_time = 200  # seconds
steps = int(total_time / dt)
c_const = np.zeros(n)
c_pulse = np.zeros(n)
surface_const = []
surface_pulse = []
high_flux = 5.0; rest_flux = 0.0; period = 20

def update(c, flux):
    new = c.copy()
    new[0] = c[0] + (D * dt / h**2) * (c[1] - c[0]) * 2 + flux * dt / h
    for i in range(1, n - 1):
        new[i] = c[i] + D * dt / h**2 * (c[i + 1] - 2 * c[i] + c[i - 1])
    new[-1] = new[-2]
    return new

for step in range(steps):
    t = step * dt
    c_const = update(c_const, flux=2.0)
    surface_const.append(c_const[0])
    flux = high_flux if (t % period) < (period / 2) else rest_flux
    c_pulse = update(c_pulse, flux=flux)
    surface_pulse.append(c_pulse[0])

time = np.arange(steps) * dt
indices = np.linspace(0, steps - 1, 200).astype(int)
plt.figure()
plt.plot(time[indices], np.array(surface_const)[indices], label='Constant current')
plt.plot(time[indices], np.array(surface_pulse)[indices], label='Pulsed current')
plt.xlabel('Time (s)'); plt.ylabel('Surface concentration (arb. units)')
plt.legend(); plt.grid(True); plt.title('Improved Diffusion Simulation')
plt.show()

# Capacity fade model
cycles = np.arange(0, 20001, 2000)
C0 = 100
C_const = C0 - (0.1 * np.sqrt(cycles) + 0.001 * cycles)
C_pulse = C0 - (0.08 * np.sqrt(cycles) + 0.0008 * cycles)
plt.figure()
plt.plot(cycles, C_const, label='Constant current')
plt.plot(cycles, C_pulse, label='Pulsed current')
plt.xlabel('Cycles'); plt.ylabel('Capacity retained (%)')
plt.legend(); plt.grid(True); plt.title('Improved Capacity Fade Model')
plt.show()
\end{verbatim}

\hypertarget{references}{%
\subsection{References}\label{references}}

{[}1{]} Ibad Ather, ``Sodium-ion batteries in 2025: a snapshot of the
fast-emerging''post-lithium'' option,'' \emph{Synergy Files}, Jun 19,
2025.

{[}2{]} W. Zhang, Y. Yang, et al., ``Carbon-bridged Nb$_2$O$_5$ mesocrystals
for fast pseudocapacitive sodium storage,'' \emph{Nanomaterials},
vol.~13, no. 5, 2023, demonstrating specific capacities of 133.4~mAh~g$^{-1}$
at 50~C and 80.5~\% capacity retention after 10~000 cycles at 20~$^{\circ}$C.

{[}3{]} S. Shi, X. Xu, et al., ``Pulse-driven internal resistance
dynamics enable dual-function lithium-ion batteries,'' \emph{Batteries},
vol.~11, no. 5, 2025, reporting a bidirectional pulse-current charging
protocol that reduces capacity fade by 30--50~\% under fast charging.

{[}4{]} Tycorun Energy, ``Blade Battery~2.0: BYD's new breakthrough,''
\emph{Tycorun Battery Blog}, Sep~2024, describing gravimetric energy
densities of 190--210~Wh~kg$^{-1}$ and 15-minute 10--80~\% charging for BYD's
next-generation Blade battery.

{[}5{]} IEST, ``Comparison of Tesla 4680 and BYD Blade battery
technology,'' \emph{IEST Energy Reports}, 2024, contrasting Tesla's
NCM811 4680 cells (\textasciitilde241~Wh~kg$^{-1}$, 1~000--2~000 cycle life)
with BYD's LFP Blade packs (\textasciitilde160~Wh~kg$^{-1}$,
\textgreater3~000 cycles).

\end{document}